\documentclass[a4paper,12pt]{article}

\usepackage[ansinew]{inputenc}
\usepackage{cite}
\usepackage{amssymb}

\usepackage[dvips]{graphicx}

\begin{document}

\title{Polarization dynamics in vertical-cavity surface emitting lasers}
\author{Thorsten Ackemann, Markus Sondermann\\ 
{\it \normalsize Institut f\"{u}r Angewandte Physik,
Westf\"alische Wilhelms--Universit\"at M\"unster,}\\ {\it
\normalsize Corrensstr.\ 2/4, D--48149 M\"unster, Germany.}\\{\it
\normalsize current e-mail: thorsten.ackemann@strath.ac.uk}}
\date {Author's last version, published as T.~Ackemann, and M.~Sondermann.
 Polarization dynamics in vertical-cavity surface emitting lasers.
 In:  O.~G.~Calderon, J.~M.~Guerra (editors): {\it Trends in Spatiotemporal Dynamics in Lasers. Instabilities,
Polarization Dynamics, and Spatial Structures}, p.~82-110. Research Signpost, Keralia (2005)}
\maketitle

\newpage

\begin{abstract}
Experiments and their interpretation on polarization dynamics and
polarization switching in vertical-cavity surface-emitting lasers
operated in the fundamental transverse mode regime are reviewed.
Important observations are switching events to a mode with the
lower unsaturated gain and the existence of elliptically polarized
dynamical transition states after the destabilization of the
low-frequency polarization mode.  The observations demonstrate the
need to consider explicitly the phase properties of the optical
field as well as nonlinear effects affecting polarization
selection above threshold. Good qualitative agreement is found
with a model which takes into account the spin degrees of freedom
of the light field as well as of the carriers (`spin-flip model'),
if the spin-flip rate is taken to be some tens of $10^9$~s$^{-1}$.
This constitutes a strong -- though indirect -- indication that
spin dependent processes are important in polarization selection
in the devices investigated.
\end{abstract}

\section{INTRODUCTION}

{\em Vertical-cavity surface-emitting lasers} (VCSELs) are a
relatively new type of semiconductor laser diodes, in which -- in
contrast to edge-emitting lasers -- the axis of the laser emission
is orthogonal to the plane of the active medium (parallel to the
epitaxial growth direction). The active zone usually consists
 of a small number of quantum wells. These are surrounded
by spacer layers which are typically only a half a wavelength
thick and the cavity is closed by high-reflectivity ($R\gtrsim
0.995$) Bragg-reflectors. Reviews of technical aspects of VCSELs
can be found in \cite{jewell91,ebeling99,iga00}.

Due to the short cavity length, VCSELs operate in a
single-longitudinal mode without any further measures. The active
zone and thus approximately also the emitted beam can be processed
to be circular. This facilitates fiber coupling tremendously in
comparison to the use of edge-emitting devices. Hence VCSELs are
the dominant type of laser in short-haul fiber communication
networks, e.g.\ local-area-networks. Further applications in
single-mode data transmission \cite{wiedenmann99b}, long-haul
communication systems with long-wavelength lasers  \cite{iga00},
spectroscopy \cite{affolderbach00,zappe00},  sensor applications
\cite{zappe00} and optical data storage \cite{thornton01} are
intensively studied for low-power and medium-power devices.
High-power devices are considered for material treatment, laser
pumping, free space communication and medical applications, e.g.\
\cite{grabherr99}.

However,  the  circular symmetry of VCSEL devices is not only
advantageous, but also a source of pronounced instabilities, since
the {\em polarization} state is no longer fixed by geometrical
constraints as in edge-emitting lasers. Even if most authors
report a rather strong pinning of the direction of the
polarization to the crystal axis in real devices
\cite{lihua94,pan93,choquette95,doorn96a} and a linearly polarized
emission at threshold, the polarization degrees of freedom degrade
the noise properties of the device
\cite{kuksenkov95,exter98b,giacomelli98} and may cause an
intriguing spontaneous flip to the orthogonal linear polarization
if the current is increased ({\em polarization switching (PS)},
\cite{pan93,choquette94,choquette95,martinregalado97a,exter98b,panajotov98}).
These phenomena  are obviously annoying in  most of the
applications mentioned above, since they are -- at least to some,
often large extent -- polarization-sensitive. This motivates a
thorough understanding of the mechanisms underlying polarization
instabilities -- notably PS -- in VCSELs as a first step to
develop control techniques.

It is rather well established by now, that one of the important
ingredients influencing the polarization state of VCSELs are {\em
linear anisotropies} stemming from the fact that real-world
devices have no ideal rotational symmetry. There are {\em
amplitude} and  {\em phase} anisotropies, i.e.\ {\em dichroism}
and {\em birefringence}. These are induced by unavoidable
mechanical stress via the elasto-optic effect
\cite{choquette94,doorn96a} and by the electro-optic effect
\cite{hendriks97}. The birefringence leads to a removal of the
frequency degeneracy of modes with different polarization and thus
to a frequency splitting. Due to the dichroism these modes
experience a different net gain (net gain = unsaturated gain -
unsaturated losses), which leads to a selection of one
polarization mode at threshold \cite{choquette94}.

Some authors found that also the polarization selection  {\em
above} threshold is determined by the linear anisotropies
\cite{choquette94,choquette95,exter98b,panajotov98}. These might
change due to thermal effects, which are inevitable during
cw-operation due to ohmic heating. The first treatment considered
the change of the gain anisotropy, if the  frequency split
polarization modes are thermally shifted across the gain line
\cite{choquette94}. Further proposals have been made
in~\cite{ryvkin99} and~\cite{panajotov00}, taking into account
also frequency dependent losses or strain effects in the quantum
well active region, respectively. Common to all these models is
that they explain PS by a current dependent change of the {\em
linear, i.e.\ unsaturated, net gain anisotropy} that becomes zero
and changes its sign at the point of the PS.

A different approach is based on the nonlinear dynamics induced by
the coupling of inversion populations with opposite spin and by
phase-amplitude coupling. The corresponding model by San Miguel et
al.\ is referred to as {\em spin-flip-model} or SFM
\cite{sanmiguel95b,martinregalado97b} (an excellent review can be
found in \cite{sanmiguel99}). We mention that phase-amplitude
coupling is particularly strong in semiconductor lasers due to the
asymmetric gain curve and often described by the so-called
linewidth enhancement or $\alpha$-factor \cite{henry82}. In the
SFM, PS occurs due to a phase instability, i.e.\ a change of the
phase relationship of the left- and right-handed circular
polarized components of the emitted light. An important prediction
of this model is that PS from the {\em mode favored by the net
gain to the gain disfavored mode} is possible.

The situation is even more complicated because it was found in a
series of careful measurements \cite{exter98b,willemsen99} that
the steady state polarization selection in many devices was
apparently determined by the linear anisotropies (and the PS due
to a change of linear dichroism) -- though the decisive mechanism
remained unclear (see, e.g., Sec.~XI of \cite{exter98b}) -- but
that dynamical features present in details of optical spectra and
transient measurements of the PS were in agreement with
predictions of the SFM. Furthermore, spatial-hole burning might
contribute to all situations \cite{valle96,martinregalado97c}.

Hence,  the question is how to differentiate among the proposed
mechanisms from an experimental point of view. One approach is to
operate the laser with current pulses with a duration far below
the thermal relaxation times. A PS from the mode with higher to
the mode with lower optical frequency (HF- to LF-mode, often
referred to as `type I PS', \cite{ryvkin99}) under pulsed
operation has been observed in \cite{martinregalado97a} and
in~\cite{verschaffelt03}. However, it was argued that this does
not give conclusive evidence for a non-thermal switching, since
the plasma temperature might change even if the lattice
temperature remains constant \cite{ryvkin99a}.

Another possibility is to look for `fingerprints' of the nonlinear
dynamics. The SFM  predicts for the PS from the LF-mode to the
HF-mode (`type II PS' in the terminology of \cite{ryvkin99}) the
occurrence of elliptically polarized dynamical states at the PS,
which lead to a reduction of the fractional polarization and
appear as sidebands in the optical spectrum. Within the framework
of the above model, characteristic spectra or similar phenomena
are not predicted for type I PS. One therefore has to look for
other manifestations of the switching to a mode with lower net
gain. We will find it in the fact that a  switching event  to the
gain disfavored mode is accompanied by a decrease of the output
power of the laser at the point of PS.

The paper is organized as follows: In the next section we will
discuss the experimental setup. Then, we will present experimental
results on polarization dynamics and polarization switching of
type 1 as well as of type 2. In Sect.~\ref{theory}, we will
compare these findings with theoretical predictions by the SFM.
The focus will be on our own results but we will try to put them
in context with the findings of other groups. Finally, a brief
outlook is given.

\section{EXPERIMENTAL SETUP}\label{setup}

The experiments have been performed on commercial gain-guided
VCSELs (Emcore Corp., Model 8085-2010) operating in the wavelength
region around 840-850~nm. We have chosen devices with an 8~$\mu$m
wide aperture which are specified as single-mode devices. Their
threshold currents vary between 3~mA and 5~mA (depending on device
and operating temperature). Higher order transverse modes usually
start to oscillate at current levels about two times the
fundamental mode threshold. We want to concentrate here on the
behavior within the fundamental mode regime, because the dynamics
is very complex and difficult to analyze, otherwise.

\begin{figure}
\centerline{
\includegraphics[width=10cm]{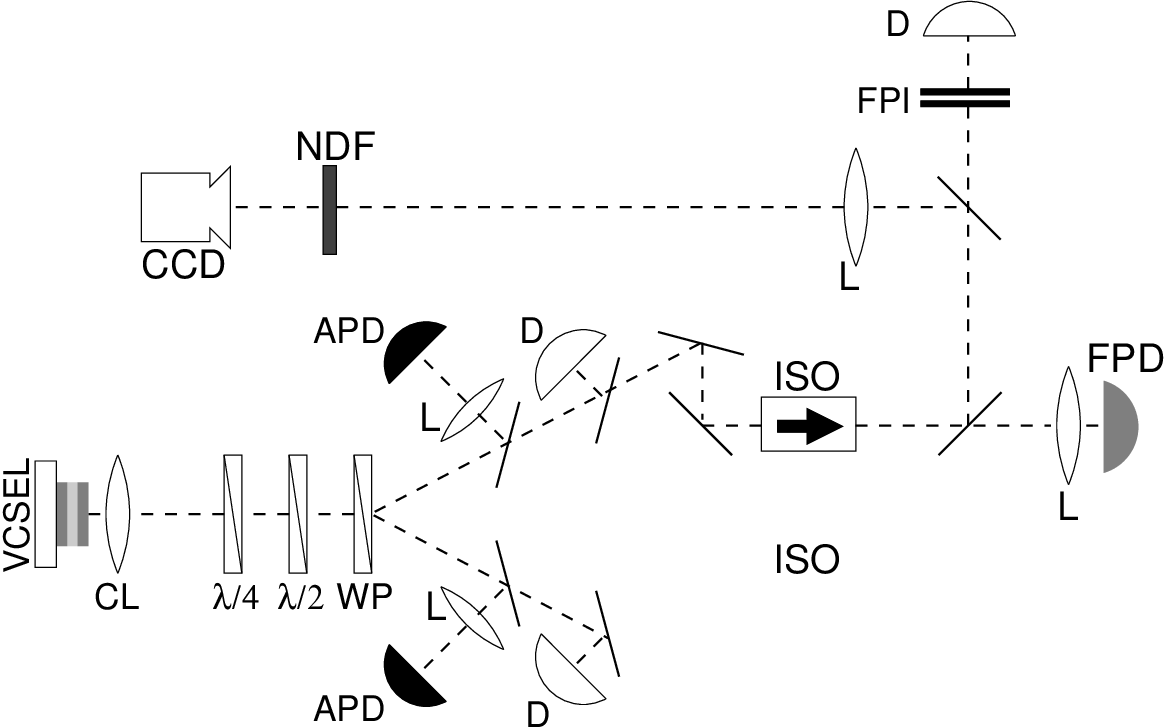}}
\caption{\label{fig:pdsetup} Scheme of the setup that was used for
the experiments on the polarization dynamics of VCSELs. The
abbreviations in the figure denote the different elements of the
setup as follows: CL, collimation lens; $\lambda/4$, quarter-wave
plate; $\lambda/2$, half-wave plate; WP, Wollaston prism; L,
lenses; APD, avalanche photo-diode (1.8~GHz bandwidth); D,
low-bandwidth detector; ISO, optical isolator; FPD, fast photo
detectors of different types (10~GHz and 26~GHz bandwidth,
respectively); FPI, scanning Fabry-Perot interferometer; NDF,
neutral density filters; CCD, charge-coupled device camera.
Mirrors   are denoted by thick solid lines. The beam path is
indicated by a dashed line. Further explanations are given in the
text. }
\end{figure}

A schematic version of the setup used to conduct the experiments
on the polarization dynamics of VCSELs is displayed in
Fig.~\ref{fig:pdsetup}. The VCSELs have been mounted in a
temperature controlled copper holder. The substrate temperature of
the VCSELs can be changed and stabilized in a range from
6$^\circ$C to 70$^\circ$C. The light emitted from the VCSEL has
been collimated using an aspheric antireflection coated lens.
After passing through a half-wave plate, the orthogonal polarized
components of the VCSEL output are split by a Wollaston prism. For
projection onto circular polarized polarization states a quarter
wave plate is inserted in the beam path in front of the half wave
plate. All polarization optics have been slightly misaligned in
order to prevent feedback into the laser. The combination of the
three elements allows the determination of the Stokes parameters,
which characterize the polarization state of the emitted light
completely. For each polarization component, the time averaged
output power and the temporal dynamics can be measured by a
low-bandwidth detector and an avalanche photo diode (APD) of
1.8~GHz bandwidth, respectively. The output of the avalanche photo
diode is recorded with a digital oscilloscope with 1~GHz analog
bandwidth (on the 5 mV/div scale). Radio-frequency (RF)
power-spectra are measured with a PIN-diode of 10~GHz or 26~GHz
bandwidth and an electrical  power spectrum analyzer of 20~GHz
bandwidth. In some measurements, time series obtained from the
26~GHz diode were directly monitored with a fast digital
oscilloscope (LeCroy Wavemaster) with 6~GHz analogue bandwidth and
a sampling rate of 20~Gs/s. Due to the low sensitivity of the
photodiode, the signal had to be amplified before by 40~dB using
two amplifiers (Agilent A83006A, 0.01-26.5~GHz) at the expense of
losing the DC-information.

A scanning Fabry-Perot interferometer (FPI) with a finesse better
than 150 and a free spectral range of 46~GHz allows for a
measurement of optical spectra. Unintended back reflections into
the laser are prevented by an optical isolator with more than
60~dB suppression. The near field intensity distribution of the
lasers was imaged on a charge-coupled devices (CCD) camera. In
experiments described here, this was used only  -- in conjunction
with the FPI -- to ensure that no high order transverse modes were
excited in the current region investigated.

It turns out that in all experiments on polarization properties of
VCSELs great care needs to be taken in the mechanical mounting of
the devices, since the linear anisotropies -- and hence the
polarization properties -- of VCSEL structures depend on
mechanical stress (e.g., \cite{doorn98,panajotov00,burak00}). On
the other hand, this sensitivity opens the opportunity to control
them. Since this possibility will be of importance later, a brief
review of the method of applying mechanical stress to VCSELs with
the intention of modifying the linear anisotropies is given here.

The elastic and elasto-optic properties of a solid can be
described by appropriate tensors. Hence, the magnitude and the
principal axes  of the birefringence can be changed by applying
external stress \cite{doorn96a,doorn98}. The latter will determine
also the orientation of the principal axis of the cavity
eigenmodes \cite{doorn96a,doorn98}. Application of stress changes
also the linear dichroism which is acting on the two polarization
modes of a VCSEL~\cite{doorn97a}, though the principal axis of the
latter one was found to be always oriented roughly along the wafer
axes for different devices and operating
conditions~\cite{doorn97a}. The orientation of the principal axes
of the birefringence  does not need to be parallel to the
orientation of the dichroism. In that case, one has to consider a
`projected dichroism', i.e.\ the dichroism that acts in the
direction of the mayor axis of the cavity mode
\cite{doorn97a,exter98b}.

The relative alignment of the anisotropies also influences the
state of polarization (SOP) of the lasing
device~\cite{doorn97a,travagnin96,travagnin97a}. As proven
experimentally in~\cite{doorn97a}, the SOP becomes elliptical if
birefringence and dichroism are misaligned. It was shown
theoretically in~\cite{travagnin97a} that modes with a finite
ellipticity are the only stable modes in this case. The
ellipticity of the SOP is measured in terms of the ellipticity
angle $\chi$. This angle is given by the arc tangent of the ratio
of the principal axes of the polarization ellipse. The second
angle that characterizes the SOP is the orientation $\phi$, which
is a measure for the tilting of the long axis of the polarization
ellipse with respect to some reference direction. As stated above,
this axis coincides with the orientation of the birefringence.

From the above considerations, it is expected that the VCSEL
emission should be always slightly elliptical  due the random
nature of the direction of stress due to contacting, bonding and
packaging. Indeed,  it was reported in~\cite{doorn97a} that all of
the investigated VCSELs had a SOP with a small but finite residual
ellipticity angle. In~\cite{exter98b} the investigated VCSELs had
an ellipticity  angle of `1$^\circ$ or less' except for the ones
with a very small birefringence (in that case it was
5-10$^\circ$). The apparent tendency of real VCSELs to emit in a
`more or less linear' polarization state oriented roughly along
the wafer axis can be explained by the rather strong anisotropy of
the tensors which relate external forces to internal strain and
then internal strain to a change in birefringence
\cite{doorn96a,doorn98}. In our setup, the `background'
contribution to the ellipticity due to imperfections in the
polarization analyzing optics is about 0.6$^\circ$. On that level,
we will call a VCSEL to be `linearly polarized'.

To summarize the above findings, the magnitude and orientation of
the birefringence, the dichroism and the ellipticity can be
manipulated by applying stress to the VCSEL. Several methods have
been proposed to achieve this purpose \cite{doorn96,panajotov00}.
The mount used follows the design introduced in
\cite{panajotov00}, where the back plate of the TO-46 package of
the VCSEL is bend over a needle providing a tensile stress on
package and wafer \cite{sondermann04t}. This method is sufficient
to change the anisotropies to the order of magnitude that is
desired, though the stress relaxes on time scales of several
hours.  It is also very difficult to match exactly the same stress
conditions in repetitive runs of the experiment. Hence, the
measurements cannot be as complete as for devices which exhibit
the desired anisotropies without application of additional stress.

Therefore, in most experiments (including all with involve a
temperature scan) a rather massive and rigid submount was used to
hold the VCSEL package in a smoothly fitting bore. This holder was
designed in order to alter the `intrinsic' anisotropies (i.e., the
ones caused by the device contacting and packaging) as little as
possible. In this mount the linear anisotropies  did not change
between runs except for small fluctuations (see also
\cite{exter98b}, Sect.~VI). On long time scales (over years) we
saw  strong changes in some cases which are attributed to aging.

For a measurement of the anisotropies  two different methods have
been established \cite{exter98b}, which both rely on the fact that
the anisotropies determine the temporal evolution of a
perturbation with orthogonal polarization to the lasing
mode~\cite{lem97,hofmann98,exter98,exter98b,mulet01a}. Note that
this definition applies also above threshold. Then `effective'
anisotropies are obtained, which are the sum of the linear (the
values at threshold) and nonlinear contributions (see also
\cite{exter98b}).

Since  a perturbation to the lasing mode, which is driven by
spontaneous emission noise, appears as the non-lasing mode in the
optical spectrum, one possibility of measuring the effective
anisotropies is an examination of optical spectra. The frequencies
and the widths of the peaks that correspond to the lasing and the
non-lasing mode are related to the effective birefringence and
dichroism, respectively, as follows: The difference in frequency
corresponds to the effective birefringence. The difference in
linewidth (HWHM) of the lasing and the non-lasing mode corresponds
to the effective dichroism~\cite{exter98b}.

A second method is the analysis of polarization fluctuations in
RF-spectra by a homodyning method. After a suitable mixing of the
polarization components on a fast detector, the central frequency
of the beating peak in the RF-spectrum corresponds to the
effective birefringence and the HWHM of the peak corresponds to
the effective dichroism \cite{hofmann98,exter98b}. Details can be
found in \cite{hofmann98,exter98b}. This method is more precise
than the method based on optical spectra since there is no
limitation set by the finite finesse of a FPI.

\section{EXPERIMENTAL RESULTS}\label{experiment}

\subsection{Polarization switching from the high-frequency to the
low-frequency mode} \label{exp_ps1}

\subsubsection{General scenario}\label{exp_general}

\begin{figure}[tb]
\centerline{\includegraphics{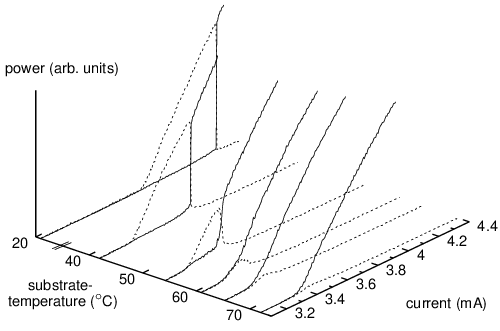}}
\caption{\label{exp_stab} Polarization resolved power against
current (LI-curve) in dependence of the substrate-temperature.
Here -- and in the following figures -- solid (dashed) lines
denote the power of the mode with lower (higher) optical
frequency. (From \cite{sondermann03a})}
\end{figure}

First, we are going to consider PS from the HF to the LF-mode.
Fig.~\ref{exp_stab} depicts the stability regions of the two
polarizations modes in the form of polarization resolved
light-current (LI) characteristics plotted for different substrate
temperature for one of the devices under study.

At about room temperature, the lasing emission at threshold is
only in the polarization direction corresponding to the HF-mode,
i.e., the lasing emission is purely linearly polarized. At
increasing current, a PS to the LF-mode is observed, as it has
been often reported in literature. If the substrate temperature of
the device is increased, the current value of the PS moves closer
to threshold. With increasing substrate temperature this
development continues until finally the point of excitation of the
LF-mode coincides with the lasing threshold (at about
60$^\circ$C). This leads to emission of both polarizations at
threshold. We will refer to this as {\em two-frequency emission}
(TFE) for reasons which will  become  clear in Sect~\ref{tfe}. At
increasing current, the mode with higher optical frequency is
depleted until only the LF-mode is lasing. If the substrate
temperature is increased even further, the emission at threshold
is dominated by the LF-mode.

This trend closely follows the temperature dependence of the
dichroism at threshold.  In the regime of excitation of both
polarizations at threshold, the absolute value of the dichroism is
less than 0.05~GHz, at 10$^\circ$C it is about 0.3~GHz. Thus, the
observation can be interpreted in the following way. For low
temperatures, the linear dichroism favors the HF-mode, for high
temperatures the LF-mode. In between there is a change of sign of
dichroism. In the vicinity of the zero, both polarization modes
are on (nearly) equal footing. Hence, both can be excited on time
average. As stated above, in general it is difficult to extract
the origin of the linear dichroism directly from the measurements.
However here, the fact that the change between the dominance of
the HF and the LF-mode occurs in or in the close vicinity of the
threshold minimum hints to the {\em gain dispersion mechanism}
proposed by Choquette et al.\ \cite{choquette94,choquette95} as
the dominant effect. The basic idea behind this is quite simple:
Since the two polarization modes are typically split in frequency
(here the birefringence is 6~GHz) they will experience a slightly
different gain. If the gain maximum has a higher frequency than
the (average) longitudinal cavity resonance, the HF-mode will be
favored, otherwise the LF-mode. If the device temperature is
changed, the detuning condition changes since the wavelength of
the gain maximum and the cavity resonance shift with quite
different rates with temperature (about 0.3 nm/K for the gain
\cite{choquette94,morgan95}, 0.08~nm/K for the cavity resonance).
In a first approximation (i.e., neglecting the decrease in gain
and the increase in relaxational processes with increasing
temperature), the position of minimal threshold for the
fundamental mode will be the temperature were gain maximum and
cavity resonance are aligned. Hence different polarization modes
can be expected to be dominant on the different sides of the
threshold minimum as seen in Fig.~\ref{exp_stab} (see also
\cite{choquette94,choquette95}).

In the following, we will discuss details of the polarization
switching at low temperature (Sect.~\ref{ps}) and of the
TFE-regime at high temperature (Sect.~\ref{tfe}).

\subsubsection{Polarization switching} \label{ps}

\begin{figure}[h!]
\centerline{
\begin{picture}(150,42)
\put(0,40){a)} \put(0,0){\includegraphics{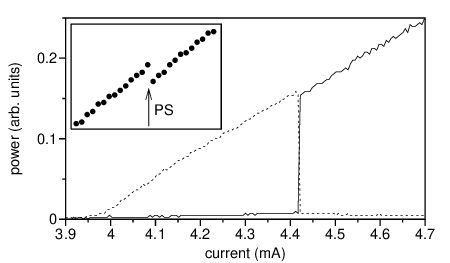}}
\put(75,40){b)}\put(75,0){\includegraphics{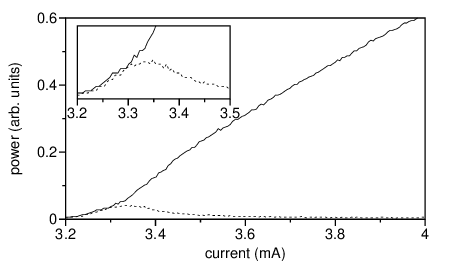}}
\end{picture}}
\caption{\label{exp_li1} Polarization resolved output power in
dependence of the injection current.  a) PS at a substrate
temperature of 10$^\circ$C. The inset displays a magnification of
the total output power (dotted line) in the vicinity of the PS,
measured without polarization sensitive optics. b) TFE at a
substrate temperature of 61$^\circ$C. The inset shows a
magnification of the current interval around threshold. (From
\cite{sondermann04b,sondermann03a})}
\end{figure}

Figure~\ref{exp_li1}a displays the polarization resolved LI-curve
under cw-operation for a device temperature of 10$^\circ$C. At
threshold the light is emitted in the HF-mode. A PS to the LF-mode
is observed at 11\% above the threshold current. At the PS a small
decrease of the output power of approximately 3\% is observed,
i.e., the LF-mode  has a lower emission power  than the
orthogonally polarized mode at the point of the PS. A linear
interpolation of the power of the LF-mode intersects the current
axis at a current value that is higher than the lasing threshold,
which is the threshold of the HF-mode. This indicates that the LF-
mode has a higher threshold. This observation is remarkable, since
lasers normally tend to choose the state with the highest output
power, if mode selection is only due to the balance of linear
(unsaturated) gain/loss anisotropies. Hence, this is an indication
that a treatment based solely on linear anisotropies is not
sufficient.

In order to ensure that the observed decrease of power is not due
to a residual anisotropy  of the analyzing polarization optics, we
have placed a low bandwidth detector directly after the
collimation lens. The total output power obtained from this
measurement exhibits a abrupt decrease at the PS in accordance
with the polarization resolved experiments (see inset in
Fig.~\ref{exp_li1}a). This is  confirmed further by a measurement
of the relaxation oscillation (RO) frequencies, which shows a
stepwise decrease at the PS point indicating a decrease in
intra-cavity power.

The scenario described up to now is observed for substrate
temperatures of the VCSEL ranging from 6$^\circ$C to 55$^\circ$C,
i.e.\ the temperature value up to which a clear, discontinuous PS
can be observed (further details can be found in
\cite{sondermann04b}). Thus, the PS is a robust phenomenon
observed in a temperature range of almost 50$^\circ$C. However,
the increase of the active zone temperature with current is only
3$^\circ$C to 4$^\circ$C per~mA and the observed PS occurs at
current values less than 1~mA above the lasing threshold, i.e.\
the temperature increase before the PS should be less than
5$^\circ$C. This is an indication that the PS observed in this
VCSEL is not due to the mechanism proposed in
\cite{choquette94,choquette95}, i.e.\ not due to the fact the
temperature of the active zone crosses the temperature for optimal
alignment of gain and cavity resonance due to Ohmic heating. This
is confirmed by the fact that a PS is also observed, if the VCSEL
is biased below threshold and driven in addition by current pulses
with a width of 10-50~ns -- shorter than the relaxation time of
the lattice temperature \cite{martinregalado97a,panajotov98} --
and a low duty cycle of 1~kHz \cite{engler03,sondermann04t}. Here
the increase of lattice temperature due to the pulses is
negligible. However, effects due to the change in plasma
temperature \cite{ryvkin99a} cannot be excluded. Polarization
switching from the HF to the LF-mode at constant lattice
temperature of the active zone were reported before
\cite{martinregalado97a,verschaffelt03}, but in these papers no
drop of output power at the PS is mentioned. The existence of drop
in output power after a PS was, however, recently reported also
for optically pumped long-wavelength VCSELs \cite{matsui03}.

\subsubsection{Two-frequency emission} \label{tfe}

Figure~\ref{exp_li1}b shows the polarization resolved LI-curve
with the  substrate temperature set to 61$^\circ$C in greater
detail. At threshold, both of the orthogonal polarization modes
start to lase with equal time averaged power. Up to approximately
4\% above threshold the power increases equally for both modes.
Then a preference for the LF-mode is observed, though the power in
the HF-mode still increases up to nearly 6\% above threshold. For
further increasing current the power in this mode continuously
decreases until it reaches the spontaneous emission level. The
time-averaged optical spectrum at threshold shows two peaks of
equal magnitude with orthogonal linear polarizations corresponding
to the two modes, i.e., the presence of power in both polarization
directions can not be attributed to a single elliptically
polarized lasing mode. Hence, we will call the current interval in
which both modes are lasing {\em two-frequency emission} (TFE)
regime. Dynamics, correlation properties and spectra of the two
modes in the TFE-regime were investigated in detail in
\cite{sondermann03a}. Here we review only the main findings.

At threshold, in both polarization directions bursts starting from
the spontaneous emission level are observed.  The amplitude of the
fluctuations is of the same order of magnitude as the average
power. The bursts have amplitudes of an equal order of magnitude
for both polarization components and appear with the same
probability in a fixed time interval. This corresponds to the fact
that the time averaged power was observed to be equal for both
polarization modes.  In the RF-spectra, relaxation oscillation
(RO) peaks are observed for both modes. This is a confirmation
that both modes are lasing. If the current is increased beyond 4\%
above the threshold value, the bursts in the HF-mode appear less
frequently than the ones in the other mode and their amplitude
decreases, until -- at high currents -- the lasing LF-mode is
fluctuating around a DC-level and the non-lasing HF-mode shows
only small-amplitude fluctuations on the spontaneous emission
level.

\begin{figure}[tb]
\centerline{\begin{picture}(150,50)
\put(0,0){\includegraphics{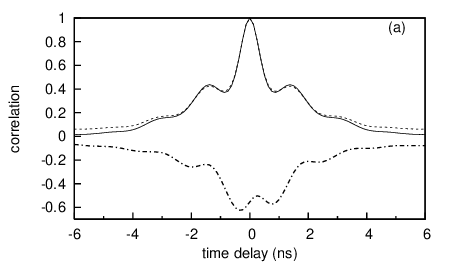}}
\put(75,0){\includegraphics{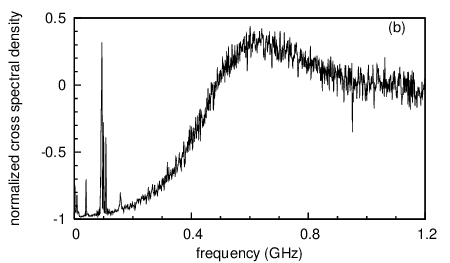}}
\end{picture}}
\caption{\label{tfe_corr} Correlation properties of the dynamics
4\% above threshold in the TFE-regime. a) The solid (dashed) line
displays the auto correlation function of the mode with lower
(higher) optical frequency, the dash-dotted line represents the
cross correlation function of the two modes. b) Normalized cross
spectral density. The fluctuations visible for frequencies less
than 150~MHz are induced by perturbations due to broadcast radio
signals. They have been checked to  enter into the APDs and to
appear also if the output of the VCSEL is blocked. (Adapted from
\cite{sondermann03a})}
\end{figure}

In Fig.~\ref{tfe_corr}a,  the  correlation properties of the time
series at 4\% above threshold in the TFE-regime are shown.  The
auto correlation functions of both modes exhibit a distinct
modulation with the frequency of the RO. This modulation is also
present in the cross correlation function of the dynamics. The
cross correlation function reveals a clear anti-correlation of the
dynamics in the two polarization directions and is slightly
asymmetric with respect to zero time lag. The (anti)correlation
decays to zero within a few nanoseconds. The absolute value
anti-correlation increases with current and reaches a value of
-0.72 at $I=3.36$~mA. Afterwards,  i.e.\ if the depletion of the
HF-mode sets in, it decreases again.

To clarify the influence of the different frequency components on
the dynamics, we have computed the normalized cross spectral
density (NCSD) of the two time traces $I_{x,y}(t)$ of the
polarization components by Fourier transformation (denoted in the
formula below by a tilde) and use of the relationship
\begin{equation}
C(f)=\mbox{Re} \left[
\frac{\tilde{I_x}(f)\cdot\tilde{I_y}^\star(f)}
{\sqrt{|\tilde{I_x}(f)|^2|\tilde{I_y}(f)|^2}} \right] \, .
\end{equation}
The NCSD contains information about the amount of
(anti)correlation at a certain frequency. The results are given in
Fig.~\ref{tfe_corr}b. At low frequencies, the dynamics are almost
perfectly anticorrelated  except for technical noise.  This
corresponds to the overall anticorrelation at zero time lag in the
cross correlation function in the time domain (see
Fig.~\ref{tfe_corr}a) and the rather slow decay of the
anticorrelation for larger time lags. Anticorrelation at low
frequencies has been shown to be a robust feature of the
polarization dynamics of VCSELs in past
investigations~\cite{willemsen99a,vey99}. For increasing
frequencies, the correlation increases (i.e., the modulus of the
anticorrelation decreases) until the NCSD reaches a maximum at
0.6~GHz with a normalized correlation value of about $0.3$. At
further increasing frequency, the NCSD decays towards zero. The
frequency of the maximum corresponds to the RO frequency of the
total power. Since the RO are a process that acts on the total
inversion and both polarization modes are lasing in the regime
under study, they have to be influenced simultaneously.

This observation explains why the anticorrelation at zero time lag
shown in Fig.~\ref{tfe_corr}a is not complete (i.e., -1) although
the NCSD at low frequencies is perfect: at zero time lag we have a
contribution from all the frequency components of the NCSD, the
high-frequency components reducing the anticorrelation due to the
low-frequency ones.

We will argue later that the TFE is the manifestation of
bistability between the two polarization state in a situation very
close to threshold where noise plays a particularly strong role
due to the weak damping of the relaxation oscillations and the
strong influence of the spontaneous emission.

\subsubsection{Some aspects of the dynamics of polarization switching}

In this section, we are going to give some additional information
on the dynamical aspect of polarization switching.

\begin{figure}[h!]
\centerline{
\begin{picture}(150,35)
\put(0,0){\includegraphics[width=75mm]{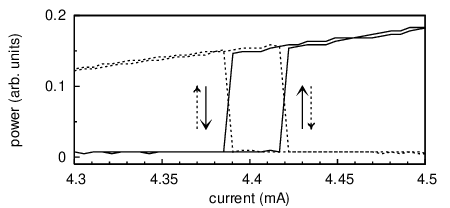}}
\put(0,32){a)}\put(75,32){b)}
\put(75,0){\includegraphics[width=75mm]{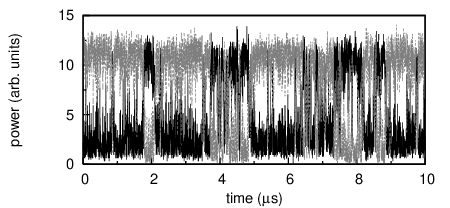}}
\end{picture}}
\caption{\label{exp_hyst} a) Polarization resolved LI-curve
including the measurement for increasing and decreasing current
(substrate temperature $10^\circ$C). The arrows indicate the
direction of the polarization switching. The scanning time for the
complete measurement is 160~seconds. b) Time series with a cw bias
within the bistable range (substrate temperature $40^\circ$C, grey
line LF-mode, black line HF-mode).}
\end{figure}

First, Fig.~\ref{exp_hyst}a shows that there is hysteresis around
the PS point, i.e., the switching points differ, if the current is
increased or decreased. This indicates that there is bistability
between the two polarization modes within the hysteresis loop.
This might open the possibility of using polarization to encode or
process information and use VCSELs for an all-optical processing
of these data (e.g.\ \cite{kawaguchi95,nieuborg98}). However, the
width of this hysteresis loop is usually quite small, only about
20~$\mu$A in this example which can be regarded as typical. We
mention that this changes drastically, if high order transverse
modes are present in the switching process. In that case
hysteresis loops can reach several mA \cite{ackemann01d}.

Since it is known that the width (or even the apparent existence)
of a hysteresis loop might depend on the ramping speed of the
stress parameter due to phenomena like critical slowing-down
(e.g.\ \cite{mandel84,mitschke83a}), the existence of bistability
was confirmed by biasing the laser at a working point within the
hysteresis loop and looking for noise-induced transitions between
the two states. Fig.~\ref{exp_hyst}b shows a time series
displaying a square-wave like hopping between the two linear
polarization states. Plotting histograms of the amplitude
distribution of each of the two polarization components yields
double-peaked (so-called `bimodal') distributions
\cite{sondermann00m} which are a proof of bistability (e.g.\
\cite{lange85} and Refs.\ therein). Since the properties of the
stochastic hopping dynamics within the bistable range are treated
intensively in the literature (e.g.\
\cite{giacomelli99,willemsen99,nagler03}), we do not go into
details here. We mention that the cross correlation function
reaches -1 at a time lag of zero and decays then slowly within a
few hundred ns towards zero. This reflects the fact that the time
scale of the competition dynamics is 100~ns to microseconds. This
longer time scale and the square-wave shape of the envelope is the
distinct difference to the dynamics in the TFE-regime. The
additional difference is that the modulations due to the ROs are
no longer pronounced since the damping of the ROs increases with
increasing distance to threshold.

Looking at a single switching event, it turns out that the
`switching time', i.e., the transition time to the new
polarization state defined by a 10\%-90\%-criterion, lies in the
range between one nanosecond and some nanoseconds depending on
parameters \cite{sondermann00m}. This is in accordance with
results from other groups \cite{willemsen00,verschaffelt00}. The
same time scale applies for a PS induced by sweeping the current
sufficiently far beyond the limit point of the bistable region. It
should be noted, however, that the observation of this short time
scale  does not allow any claim on a non-thermal origin of the
switching, because the thermal contribution might be hidden in a
long lethargic stage in which the system is very close to the
original state. This initial stage is not captured by a
10\%-90\%-criterion.  Only modulations experiments
\cite{verschaffelt03} or pulsed excitation (see above) can help to
distinguish between thermal and non-thermal origins of the PS.

\subsection{Polarization switching from the low-frequency to the
high-frequency mode} \label{exp_ps2}

Apart from the PS type 1 discussed in last section, switching from
the LF to the HF-mode (so-called `PS of type 2') is described also
in the literature
\cite{exter98b,ryvkin99,willemsen00,ackemann01c,sondermann04a}. It
is instructive to compare the properties of type 1 and type 2
switching since the SFM allows for both, but gives definite
different predictions for the two cases \cite{martinregalado97b}.

\begin{figure}[htb]
\begin{center}
\begin{picture}(150,35)
\put(0,0){\includegraphics{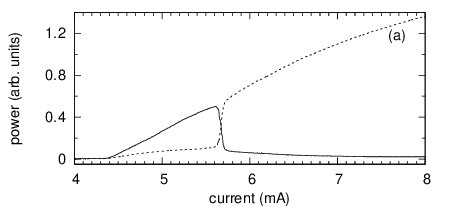}}
\put(75,0){\includegraphics{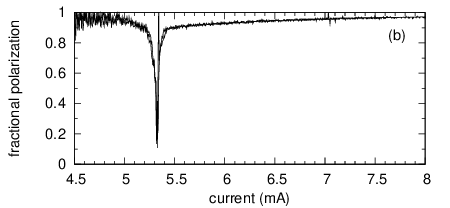} }
\end{picture}
\end{center}
 \caption[]{\label{exp_li2}
 a) LI-curve after projection on
 linear polarization states with the axes parallel to the principal
 polarization axes at the PS. Substrate-temperature 15.9$^\circ$C.
 b) Fractional polarization. The vertical line denotes the PS point.
 The strong noise for low current is due to the fact that the signal level is small.
  Substrate temperature: 22.6$^\circ$C. (Adapted from
  \cite{ackemann01c,ackemann01d})}
\end{figure}

Fig.~\ref{exp_li2}a shows a LI-characteristic with a PS of type 2
for one of our devices. Interestingly, for projection onto
orthogonal linear polarization components there is always also
significant excitation of the weaker polarization. This is due to
the fact that the lasing mode is in fact elliptically polarized,
as it can be shown by a measurement of the Stokes parameter,
(ellipticity angle at threshold 10.5$^\circ$, see Fig.~7a of
\cite{ackemann01d}). Far enough away from the PS, the  {\em
fractional polarization}, i.e.\ the sum of the squares of the
normalized Stokes parameters, is nearly one (Fig.~\ref{exp_li2}b),
i.e.\ there is a still a pure -- though elliptical -- SOP. Another
difference to the scenario discussed in the preceding section is
that the fractional polarization drops to rather low values in a
rather broad interval around the PS point. This indicates that
there is no pure polarization state in this regime. Indeed, in the
optical spectrum there are multiple lasing lines with elliptical
polarization but different principal axes (see Fig.~2 of
\cite{ackemann01c}). This indicates that the emission of the laser
is not constant but that there is a {\em dynamical, self-pulsing
state}. We will discuss details below. We never observed such a
behavior  for a PS of type~1.

This behavior is rather robust. The LF-mode is the mode selected
at threshold for the whole temperature range investigated
($10^\circ-65^\circ$C, see Fig.~5 of \cite{sondermann04a})). At
high currents, the HF-mode is active. In between, there is the PS
with the transition region with dynamical states. The region with
dynamical states is sickle-shaped and closes, in tendency, for
increasing temperature.

In the laser, in which the SOP is elliptically at threshold, the
principal axes of the SOP are not aligned to the wafer axes and
rotate continuously with current by a rather large amount, about
15$^\circ$ (see Fig.~7a of \cite{ackemann01d}; the principal axis
of a `linear' (see Sect.~\ref{setup}) SOP are fixed within
1$^\circ$ or less). This hints to the fact that the principal axis
of birefringence and dichroism are not aligned in these devices as
discussed in Sect.~\ref{setup}.  For a comparison between theory
and experiment as well as between PS of type 1 and PS of type 2,
it is an important question, whether the dynamical states can also
be observed experimentally for other parameter values, i.e., for
other values of linear dichroism and birefringence and especially
for lower values of the ellipticity angle of the steady state at
threshold.  To induce the parameter changes, stress was applied to
different devices with the method explained in Sect.~\ref{setup}.
With this method the linear anisotropies were changed until a PS
from the LF-mode to the HF-mode occurred. The results are
summarized in \cite{sondermann04a}.

\begin{figure}[h!]
\centerline{\includegraphics{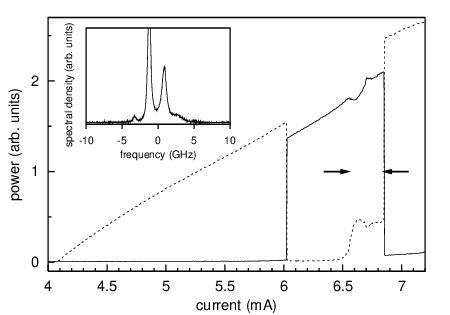}}
\caption{\label{exp_double} Polarization resolved output power
against current for projection onto the main axes of the state of
polarization at threshold for a device with an ellipticity angle
$\le$1$^\circ$ at threshold. The inset shows the optical spectrum
for a current of 6.66~mA after projection onto linear
polarization. The largest peak is cut off for a better
visualization of the smaller peaks. The arrows denote the current
region in which sidebands  are observed in the optical spectrum.
(From \cite{sondermann04a})}
\end{figure}

Here, we will discuss in detail a particularly interesting
scenario which is depicted in Fig.~\ref{exp_double}. At threshold,
the lasing mode is the HF-mode. The SOP can be regarded as
linearly polarized, the ellipticity angle is smaller than
1$^\circ$ (see the discussion at the end of Sect.~\ref{setup}). At
increasing current, at first a PS to the LF-mode occurs (note the
drop in power). Up to a current of 6.5~mA -- i.e., in a current
range extending beyond the first PS -- the ellipticity stays below
4$^\circ$. Also dynamical states do not appear at the first PS. If
the current is increased further (beyond 6.5~mA), the ellipticity
angle strongly increases (see also the increase of power in the
weak linear polarization component about 6.4~mA) and reaches about
22$^\circ$ before the second PS. At a current value of 6.55~mA
sidebands appear in the optical spectrum and the fractional
polarization decreases strongly (the minimum value is about~0.6).
The sidebands disappear at the second PS, which is a switching
back to the HF-mode. The fractional polarization increases again
after the PS and the ellipticity drastically decreases. This
scenario reported here is qualitatively similar to the one
observed for a single PS from the LF- to the HF-mode
\cite{sondermann04a}.

The results obtained so far indicate that the existence of
elliptically polarized dynamical states before a  PS of type 2
does not depend qualitatively on the initial ellipticity of the
SOP at threshold, i.e., the scenario for the destabilization of
the LF-mode is: LF-mode lasing (nearly linearly polarized) $\to$
increase of ellipticity $\to$ appearance of multiple peaks in the
optical spectrum and drop of fractional polarization $\to$ PS to
HF-mode. As we will see later, these observations confirm the core
of the predictions of \cite{martinregalado97b} for the transition
scenario.

\begin{figure}
\centerline{\includegraphics{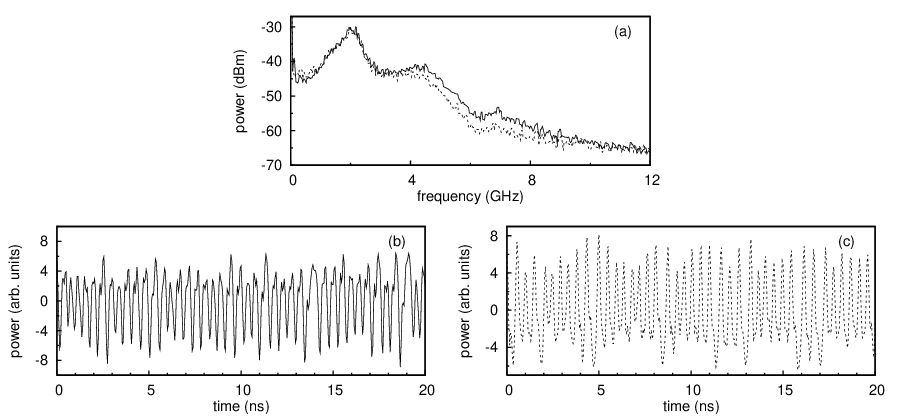}}
\caption{\label{exp_dynamics} Power spectra (a) and time series
(b,c) of the dynamics in the regime of the emission of dynamical
states. The solid (dashed) line in (a) represents the power
spectrum after projection onto the linearly polarized state with
maximum (minimum) dc-output. The corresponding time trace is given
in (b) and (c), respectively. All data were measured after 40~dB
amplification. Note that the two time traces have not been
obtained simultaneously and that the DC-information is lost due to
the amplification. (Adapted from \cite{sondermann04a})}
\end{figure}

In some of the experiments a fast digital oscilloscope was
available, allowing for an investigation of the temporal dynamics
during the emission of the dynamical states. In
Fig.~\ref{exp_dynamics} the dynamics is illustrated for the same
device that is presented in Fig.~\ref{exp_li2} but for a different
run of the experiment.  As depicted in Fig.~\ref{exp_dynamics}(a),
a strong peak at a frequency of about 2.1~GHz is observed in the
power spectrum for projection onto the linear polarization
directions corresponding to maximum and minimum time averaged
power, i.e., onto the main axes of the polarization ellipse of the
time averaged SOP. This frequency corresponds to the distance of
the sidebands in the optical spectrum (in optical frequencies) and
can therefore be interpreted as the beating frequency between the
different `modes' which are oscillating at adjacent frequencies.
We interpret this frequency difference as the `effective
birefringence', i.e., the sum of the linear birefringence observed
at threshold and the non-linear contributions due to saturable
dispersion and spin dynamics
\cite{lem97,hofmann98,exter98a,exter98b,ackemann01c}. Since
several sidebands are excited in the optical spectrum (see also
inset of Fig.~\ref{exp_li2}), also higher harmonics of the beating
frequency are observed in the power spectrum.

Corresponding to these frequencies, pronounced oscillations are
observed in the time domain (see Fig.~\ref{exp_dynamics}(b) and
(c)). The time traces have not been obtained in a simultaneous
measurement. Hence they contain no information about the
correlation properties of the dynamics. The oscillation at 2.1~GHz
is observed for projection onto both of the polarization main
axes.

From these observations the following conclusions can be drawn:
First, due to the rather regular temporal oscillation at the
frequency splitting of the sidebands in the optical spectrum, the
sidebands can be considered as a (nearly) locked state and not as
independent modes. Second, the presence of these oscillations in
both linear polarization components hints to an oscillation of the
characteristic polarization angles, i.e., ellipticity and
orientation of the SOP. This issue is discussed further in the
theoretical section.

\section{THEORETICAL INVESTIGATIONS AND DISCUSSION}\label{theory}

\subsection{Model equations}\label{model}

The experimental observation clearly indicate two important
features any model should be able to explain (apart, of course,
from the obvious fact that there is polarization switching):
\begin{itemize}\itemsep0cm
\item The ellipticity of the optical field was observed to vary.
Hence, one has to take into account the {\em relative phase}
between the two polarization components and thus needs to treat
equations for the complex optical fields, not only the intensities
as often done in the literature.
\item The model should include the possibility that switching
occurs to the mode with the lower unsaturated gain, it should not
concentrate on a change of the linear gain-loss balance as the
basic mechanism behind polarization switching.
\end{itemize}
To our knowledge, the spin-flip model
\cite{sanmiguel95b,martinregalado97b} is the only model worked out
in the literature which is capable of fulfilling these
requirements. Especially, dynamical states have not been reported
in the framework of other models until now.  In particular, we
will use an  extended version of the spin-flip model that includes
a realistic semiconductor
susceptibility~\cite{balle99,sanmiguel00a}. Thanks to the
frequency dependence of the susceptibility one is capable to
correctly describe changes in the relative position between the
cavity resonance and the gain curve. Hence, the thermal shift of
resonances due to temperature changes, which is important, e.g.\,
for an explanation of the observations depicted in
Fig.~\ref{exp_stab}, can be easily taken into account. It will
turn out that it is possible to explain all the observations
presented before in a single model using essentially the same
parameters except for the linear anisotropies, which obviously
have to be adjusted to any new situation
\cite{sondermann03a,sondermann04a,sondermann04b}. This does not
exclude that it is possible -- and interesting -- to describe
important aspects of a particular phenomenon by a reduced set of
equations derived from the SFM (e.g.\ \cite{vandersande03} for PS
of type 1 and \cite{travagnin97} for PS of type 2).

Under fundamental transverse mode operation, the evolution of the
circularly polarized components of the electric field $E_\pm$
(slowly-varying envelopes) and the electronic densities $D_\pm$
with opposite spin (normalized to the transparency density $N_t$)
are governed by~\cite{balle99,sanmiguel00a}:
\begin{eqnarray}
\label{Epm} \dot E_{\pm}(t)& = &-\kappa E_{\pm}+i\frac{a\Gamma}{2}
\ \chi _\pm\!\left(\Omega+i\frac{\dot
E_{\pm}}{E_\pm},D_+,D_-\right)E_{\pm}
-(\gamma _a+i\gamma _p)E_\mp+\sqrt{\beta _{sp}D_\pm}\xi
_\pm(t) ,\\
\label{Dpm} \dot D_\pm (t)& = &\frac{1}{2} \mu \frac{I_t}{e N_t}
-AD_\pm -BD_\pm^2 \mp\gamma_j(D_+-D_-) 
+a\cdot Im\chi_\pm\!\left(\Omega+i\frac{\dot
E_{\pm}}{E_\pm},D_\pm\right)|E_\pm |^2 .
\end{eqnarray}
The electronic densities with opposite spin interact with
circularly polarized light with different helicity through the
frequency dependent susceptibility $\chi_\pm$
\cite{balle99,balle98}. We use analytical expressions for
$\chi_\pm$ obtained in~\cite{balle98} that describe the gain and
refractive index spectra as
\begin{eqnarray}
\label{susc}
\chi _\pm(\omega_\pm+\Omega,D_\pm) 
-\chi_0 \left[\mathrm{ln}\left( 1-\frac{2D_\pm}{u_\pm+i}\right)
+\mathrm{ln}\left(1-\frac{D_+ + D_-}{u_\pm+i}\right)
-\mathrm{ln}\left(1-\frac{b}{u_\pm +i}\right)\right] 
\end{eqnarray}
where
\begin{equation}
\label{upm} u_\pm = \frac{\omega_\pm}{\gamma_\bot}+\Delta+\sigma
(D_+ +D_-)^{1/3}\quad , \quad
\Delta=\frac{\Omega-\omega_t}{\gamma_\bot}\quad .
\end{equation}
$\Delta$ is the detuning between the cavity resonance $\Omega$ and
the nominal transition frequency $\omega_t$ of the band gap,
normalized to the material polarization decay rate $\gamma_\bot$.
Thus, the difference in thermal shift of the frequency of the gain
maximum and of the cavity resonance can be modeled by a variation
of $\Delta$. $\Delta$ increases with temperature since the
redshift of the cavity resonance increases faster with temperature
than the redshift of the band gap frequency~\cite{choquette94}.
$\sigma$ describes band-gap shrinkage and $b$  is a background
contribution to $\chi$ without pumping. Spin-flip processes that
reverse the electron spin directly couple the two carrier
reservoirs. This effect is phenomenologically accounted for by
means of the spin-flip rate $\gamma_j$. $\mu$ is the injection
current normalized to the transparency current ($I_t$), $e$ is the
electron charge. The linear contributions to the birefringence and
dichroism are $\gamma_p$ and $\gamma_a$, respectively. In the
framework of this model $\gamma_a$ is a pure loss anisotropy. The
differences in material gain due to the frequency splitting
between the modes are incorporated by the optical susceptibility
$\chi_\pm$. The rest of parameters are the cavity losses $\kappa$,
the effective gain constant $a$, the confinement factor $\Gamma$,
the non-radiative and bimolecular recombination rates of the
carriers $A$ and $B$ and the spontaneous emission rate
$\beta_{sp}$. Finally, $\xi_\pm(t)$ are white Gaussian random
numbers with zero mean and delta correlated in time that model
spontaneous emission processes.

For convenience, the simulations have been performed in the
circular polarized basis. To obtain expressions for the linearly
polarized components one has to use the relations
$E_{||}=(E_++E_-)/\sqrt{2}$, $E_\bot=(E_+-E_-)/(i\sqrt{2})$, where
steady state solutions of $E_\pm$ are of the form $E_\pm(t)=Q_\pm
e^{-i(\omega_\pm t \pm\psi )}$ with $Q_\pm$ being the amplitude of
the field and $\psi$ the phase with which the circularly polarized
fields lock. If $\gamma_p > 0$, $E_\bot$ corresponds to the
LF-mode and $E_{||}$ corresponds to the HF-mode. The parameters
used for the simulations are given in the figure captions and can
be considered to be typical VCSEL parameters. The spin flip rate
$\gamma_j$ is taken as a fit parameter. It is found that the
experimentally observed dynamics can be reproduced, if a rather
low value of some tens of $10^9$~s$^{-1}$ is assumed. We will
comment on this below.

\subsection{Theoretical investigations on PS of type 1}\label{theory_ps1}

\begin{figure}[tb]
\centerline{
\begin{picture}(150,50)
\put(0,48){a)} \put(75,48){b)}
 \put(0,0){\includegraphics[width=75mm]{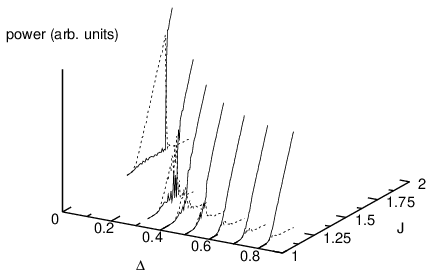}}
\put(75,0){\includegraphics{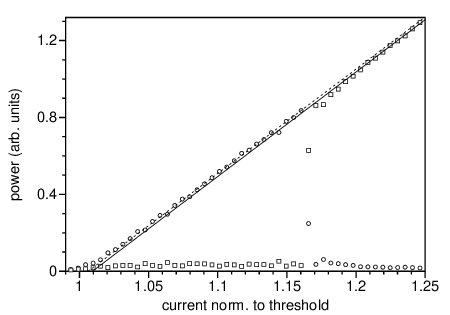}}
\end{picture}}
\caption{\label{sim_stab} a) Polarization resolved output power in
dependence of the detuning $\Delta$ and injection current $J$
normalized to its transparency value.  The parameters defining the
susceptibility function~\cite{balle98} are $b=10^4$,
$\gamma_\perp= 10^4$~ns$^{-1}$, and $\sigma=0.2$, other parameters
$\gamma_p=18$~ns$^{-1}$, $\gamma_a = 0$, $\kappa=300$~ns$^{-1}$,
$a=2.3\cdot 10^4$~ns$^{-1}$, $\Gamma=0.045$,  $A=0.5$~ns$^{-1}$,
$B=1.0$~ns$^{-1}$, $\beta_{sp}=10^{-6}$~ns$^{-1}$,
$\gamma_j=20$~ns$^{-1}$. b) Polarization resolved LI-curve
obtained from simulations for $\Delta=0$. Circles (squares) denote
the power in the mode with higher (lower) optical frequency. The
dashed (solid) lines are linear fits applied to the data before
(after) the PS. The integration time per point is 100~ns. (From
\cite{sondermann03a,sondermann04b})}
\end{figure}

First, we are going to consider the case of the destabilization of
the HF-mode. As a first step, LI-curves for various detunings have
been simulated (see Fig.~\ref{sim_stab}a) assuming that there is
no loss or strain induced dichroism, $\gamma_a=0$. By a change of
the detuning a variation of substrate temperature as performed in
the experiments can be simulated.  As it is obvious from
Figs.~\ref{exp_stab} and \ref{sim_stab}a, the simulations
reproduce the experimentally measured polarization properties
qualitatively for variation of both the detuning {\em and} the
current. The selection of the lasing mode at threshold is, of
course, the result of the gain dispersion mechanism proposed by
Choquette et al.\ \cite{choquette94,choquette95} and discussed in
the experimental section.

In the discussion, we concentrate first on the situation for small
detuning (corresponding to low temperature in the experiment). In
Fig.~\ref{sim_stab}b the simulated LI-curve for a detuning value
of $\Delta=0$ is displayed. At threshold the laser oscillates in
the HF-mode, which is the one closer to the gain peak. If the
current is increased beyond a value of 16\% above the lasing
threshold, a PS to the LF-mode  is observed. Linear fits have been
applied to the data before and after the PS (see lines in
Fig.~\ref{sim_stab}b). The fitting results reveal a drop of the
output power at the PS and a higher threshold for the LF-mode,
i.e.\ the PS is to the mode with lower gain. The decrease of power
at the PS is of the order of 2\%. In the simulations, the linear
anisotropies were assumed to be unchanged during the current scan,
but we can also include a current-dependent drift in $\Delta$ (and
hence the gain anisotropy) in order to model the effects of Joule
heating. As long as the increase of $\Delta$ with current is
moderate the drop of the output power at the PS is still observed
\cite{sondermann04b}.

In the theory, we are in the favorable situation that there is no
doubt on the interpretation of this power drop:  The linearly
polarized steady state solutions of the equations
(\ref{Epm}),(\ref{Dpm})  are functions of $\gamma_\mathrm{a}$ and
the imaginary part of the susceptibility. This results in general
in different thresholds for the two polarization modes and in
different amplitudes of the linearly polarized fields at constant
current~\cite{sanmiguel00a}. Hence, the drop is a direct
consequence of the fact that the mode with the higher unsaturated
net gain becomes unstable in favor of the mode with the lower
unsaturated net gain. Though this might sound counter-intuitive on
the first sight, it is necessary to recall that the unsaturated
net gain governs only the competition for growth from the
non-lasing, zero solution, but not -- at least not necessarily --
from a lasing solution (see also \cite{burak00}). The
destabilization of the higher gain mode is due to a complex
interplay of birefringence, phase-amplitude coupling and
spin-flips \cite{sanmiguel99}. This is, e.g., apparent from the
contributions entering the cross-coupling coefficients  between
polarization modes coefficient in perturbative treatments
\cite{lem97,exter98a,vandersande03}. However, no simple
explanation seems to be available in the literature, though some
insight can be gained by considering the trajectories of
perturbations on the Poincare sphere \cite{exter98a}. We mention
that similar effects are at work in {\em detuned gas lasers}
(e.g.\ \cite{haeringen67,lenstra80}), the detuning being
responsible for phase-amplitude coupling. Since in semiconductor
lasers phase-amplitude coupling is particularly strong and, in
addition, exists even for operation at the gain peak
\cite{henry82}, the existence of these effects in VCSELs is not
totally surprising.

A detailed comparison of the hysteresis properties and the
development of the effective dichroism in experiment and theory is
contained in \cite{sondermann04b}. Also there, the experimental
results are in good qualitative agreement with the predictions of
the SFM. The power drop at the PS is also contained in versions of
the SFM which contain gain saturation (\cite{prati02}, often
called also `gain compression'), or reduced versions of the SFM in
which the processes of spin-flips and phase-amplitude coupling are
`condensed' in a cross-saturation coefficient
\cite{vandersande03}. It is also discussed in \cite{sondermann04b}
that gain saturation alone \cite{ryvkin03,danckaert02}-- i.e.\ for
zero linear dichroism -- might explain bistability, but not the
power drop. We also stress that the observation of a minimum of
dichroism or mode suppression ratio \cite{exter98b,verschaffelt03}
is -- taken alone -- not sufficient to draw conclusions on the
origin of the PS, i.e.\ whether it is solely induced by a change
in linear dichroism or not \cite{sondermann04b}.

In the TFE-regime, the important features of the dynamics, i.e.,
the overall shape of the time series, spectral and correlation
properties, can be also reproduced nicely by the SFM
\cite{sondermann03a}. The analysis shows -- as expected
intuitively -- that the PS point moves to lower current values
with increasing detuning (cf.\ Fig.~\ref{sim_stab}a)) due to the
reduction of the linear dichroism. In the threshold minimum, the
two linear polarization modes are bistable already at threshold.
In that situation, the apparent competition dynamics is
particularly strong and takes place on short time scales, since
close to threshold the influence of spontaneous emission noise is
particularly strong and the relaxation oscillations are only
weakly damped. Hence, TFE can be regarded as the manifestation of
bistability close to threshold. Although the TFE-regime cannot
provide a clear-cut distinction between the SFM and other models,
since the correlation properties between polarization modes should
resemble qualitatively the ones depicted in Fig.~\ref{tfe_corr} in
any model allowing for polarizing competition, the comparison
between SFM and experiment is favorable even in a
semi-quantitative way \cite{sondermann03a}. This is the more
remarkable since a single set of parameters was used to describe
the PS and the TFE dynamics. Finally, we mention that the
depletion of the HF-mode further above threshold is due to
nonlinear contributions to the dichroism \cite{sondermann03a}.

\subsection{Theoretical investigations on PS of type 2}\label{theory_ps2}

As mentioned above, the SFM makes a definite prediction for the
destabilization of the LF-mode in a broad parameter regime
\cite{martinregalado97b}: The linearly polarized LF-mode becomes
unstable in favor of an elliptically polarized modes at some
point, if the current is increased above threshold. At a somehow
larger current, the elliptically polarized modes becomes unstable,
too, and regular oscillations develop. If the current is increased
further, these oscillations become irregular, possibly chaotic,
until finally the system switches to the linearly polarized
HF-mode. Details of this scenario are investigated
semi-analytically in \cite{erneux99,prati04}. Due to the
complexity of the extended SFM used here, it is more difficult to
determine the elliptically polarized solutions and their
stability. Hence we performed only numerical simulations.

\begin{figure}
\centerline{\includegraphics{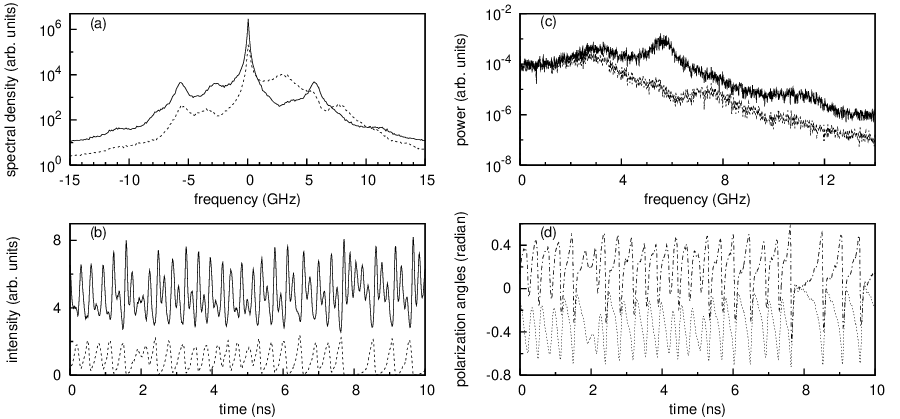}}
\caption{\label{dynsim} Spectra and temporal evolution for
projection onto linearly polarized states during the emission of
the dynamical states. The optical spectra are shown in (a). The
temporal evolution of the corresponding field intensities is
displayed in (b), whereas in panel (c) the power spectrum
calculated from the time traces (total integration time 2~$\mu$s)
is shown. The temporal evolution of the polarization angles is
given in (d), where a dotted (dash-dotted) line represents the
ellipticity (orientation) angle of the state of polarization. The
parameters are $\gamma_a$=1~ns$^{-1}$, $\gamma_p$=9~ns$^{-1}$,
$\gamma_j$=20~ns$^{-1}$, $\Delta$=0, $A$=0.5~ns$^{-1}$,
$B$=1~ns$^{-1}$, $\kappa$=300~ns$^{-1}$, $a=2.5\times
10^4$~ns$^{-1}$, $\gamma_\bot=10^4$~ns$^{-1}$,
$\beta_{sp}$=10$^{-6}$~ns$^{-1}$. The current is 1.9 times the
threshold value. (Adapted from \cite{sondermann04a})}
\end{figure}

Details can be found in \cite{sondermann04a}. LI-curves and the
existence regions of dynamical states are found to be in
qualitative agreement with experimental observation. Here, we will
discuss only the time series in the presence of dynamical states.
Typical results are depicted in Fig.~\ref{dynsim}.

Panel (a) displays the optical spectra after projection onto
linearly polarized states, which exhibits several sidemodes that
are typical for this dynamical regime. This matches the
experimental observation (see Fig.~\ref{exp_dynamics}) except for
the amplitudes of the sidebands, which are considerably more
excited in the experiment. The time traces of the corresponding
intensities are given in panel (b) and the power spectra in panel
(c) of Fig.~\ref{dynsim}, respectively. As in the experiments, the
power spectra exhibit pronounced components at the beating
frequency of adjacent side modes and its multiples. The
corresponding oscillations are observed in the time traces. The
computation of a cross correlation function reveals that the
intensities in the orthogonal linear polarization directions have
a correlation of -0.76  at zero time delay, i.e., they are
strongly anticorrelated. As in the experiment, the oscillations
are rather -- but not totally -- regular.

Unlike in the experiment, in the simulations the time resolved
ellipticity and orientation angle can be easily computed from the
time series of the complex optical field components. The result of
this procedure is displayed in Fig.~\ref{dynsim}d. Both
polarization angles are strongly oscillating. Their oscillation is
also strongly anticorrelated. The fractional polarization has a
value of 0.45. Thus these oscillations can be interpreted as an
oscillation on the Poincare sphere (see also
\cite{martinregalado97b,exter98a}).

Obviously, the destabilization scenarios are strikingly different
for the HF-mode and the LF-mode. In our opinion, the asymmetry is
due to the different mechanism of destabilization of the lasing
mode. The linearly polarized steady state is characterized by
equality of $D_+$ and $D_-$~\cite{martinregalado97b,sanmiguel00a},
which is valid for both the LF- and the HF-mode. For linearly
polarized states, the stability of the total intensity decouples
from the stability of the polarization subset. The stability of
the linearly polarized modes against perturbations with orthogonal
polarization and/or a deviation from $D_+=D_-$ is expressed by the
three remaining eigenvalues. Two of these are complex conjugate,
the third eigenvalue is real. The real eigenvalue is (at least in
the case of relatively large spin relaxation rates) mainly
associated with perturbations of the equality of $D_+$ and
$D_-$~\cite{lem97,exter98a}. In the case of switching from the HF-
to the LF-mode and for the parameter combinations we have studied,
it is always the complex conjugate eigenvalue that acquires a
positive real part and hence leads to destabilization of the
HF-mode (see also~\cite{martinregalado97b}). The real eigenvalue
is always negative, what implies stability of the condition
$D_+=D_-$. Elliptically polarized steady states are characterized
by $D_+\ne D_-$~\cite{martinregalado97b}. Hence, elliptically
polarized states cannot be expected.

For the PS from the LF- to the HF-mode, the behavior of the
eigenvalues is different. Here it is the real eigenvalue which
becomes unstable, possibly after the imaginary part of the complex
conjugate eigenvalues became before zero at increasing current.
This explains somehow the tendency to form an elliptically
polarized state with $D_+\ne D_-$.

Looking more on physical mechanisms (than eigenvalue analysis), an
asymmetry between the LF- and HF-mode exists in the framework of
the SFM, since the  nonlasing mode experience a redshift above
threshold related to phase-amplitude coupling
\cite{lem97,exter98b}. Hence, the effective birefringence
increases, if the HF-mode is lasing, as observed experimentally in
\cite{exter98b}. On the contrary, if the LF-mode is lasing, the
splitting decreases (in tendency; the scenarios can be very
involved depending on parameters, especially the spin-flip rate),
which was indeed observed experimentally in our group
\cite{ackemann01c}. It appears to be somehow intuitive that a
decrease of the birefringent splitting favors the stable `locking'
to an elliptically polarized mode.

\subsection{Influence of the spin-flip rate}

The experiments and the simulations show  good agreement for type
1 as well as type 2 switching, if the spin flip rate is assumed to
be of the order of some tens of $10^9$~s$^{-1}$.  Previous
estimations of the spin flip rate under lasing conditions were
inferred indirectly from experiments assuming the validity of the
SFM and yielded $\gamma_j \approx 30\ldots 75\cdot 10^9$~s$^{-1}$
\cite{doorn97}, $\gamma_j > 100\cdot 10^9$~s$^{-1}$
\cite{exter98b},  $\gamma_j \approx  10^{12}$~s$^{-1}$
\cite{blansett01} and `infinity' \cite{besnard99}, i.e., the
values reported in the literature span over a rather wide range
but are significantly higher,  in tendency, than the values
considered here. In that case, type I switching is indeed only
possible, if a current dependence of the linear net gain is
assumed. Also the current for PS of type 2 increases until it
cannot be reached any more at a current value reasonable for real
devices.

On the other hand, pump-probe experiments on passive quantum well
structures yielded even lower values of $\gamma_j \approx 7\ldots
9\cdot 10^9$~s$^{-1}$, even at room temperature
\cite{cameron96,ohno99}. However, it is often argued that it is
doubtful whether these values provide a good estimation for the
situation of a high carrier density typical for laser operating
conditions. Nevertheless, at least one investigation demonstrates
nearly circularly polarized lasing emission in a VCSEL after
optical pumping with circularly polarized light \cite{ando98}.
Even though the data indeed demonstrate that the spin-flip rate
increases with increasing carrier density, the  lifetime of spin
polarization at the laser threshold ($N_c\approx 3-4\times
10^{18}$~cm$^{\-3}$) was found to be 40~ps, i.e., compatible with
our assumptions. Hence we conclude that the assumptions made here
on the spin-flip rate are well within the established limits.

The very different conclusions drawn in the literature on the
relevance of spin-flip processes for polarization selection in
VCSELs are probably related to the fact that the spin-flip rate
depends on sample quality via the density of scattering centers 
\cite{britton98}. Hence we regard our results as complementing --
and not contradicting -- the earlier investigations.

\section{SUMMARY AND OUTLOOK}

In this paper, we presented experimental results on polarization
dynamics in VCSELs and their interpretation. The results,
especially the observation of self-pulsing, elliptically polarized
emission states and the switching to a gain disfavored mode,
indicate clearly the need for taking into account phase degrees of
freedom and nonlinear effects going beyond the normal `winner take
all' dynamics of the mode with the largest unsaturated, linear
gain. Good agreement was found with the spin-flip model assuming a
rather low spin-flip rate of some tens of $10^9$~s$^{-1}$. This is
taken as a strong -- though indirect -- indication that spin
dependent processes  contribute to the polarization selection in
the devices under study. Further insight is probably difficult to
achieve with ready-made commercial devices but depends on the
availability of wafer samples on which first-hand growth
characterization, pump-probe characterization of the active zone
in single pass, lasing properties with polarized optical pumping
and lasing properties with electrical pumping are accessible at
the same time. The results might be awarding in terms of {\em
spintronics} and all-optical polarization based processing
devices, even if for polarization control an enhancement of the
linear dichroism might be the simplest option in the end.

Particularly interesting is also the interplay of polarization and
transverse mode dynamics. VCSELs are also very vulnerable to the
excitation of high-order modes, because typically the Fresnel
number of the cavity is rather large. The appearance of high-order
transverse  modes is often accompanied by polarization effects
(e.g.\
\cite{chang91,lihua94,choquette95,epler96,giacomelli98,ackemann01d,barchanski03}).
Some preliminary results on the coupling of spatial and
polarization degrees of freedom are found in \cite{ackemann01d}.
Extensions of the SFM to a system of nonlinear partial
differential equations for the study of transverse multi-mode
dynamics are available \cite{mulet02,mulet02a}. The polarization
properties of spontaneously formed spatial patterns in broad-area
VCSELs are also intriguing
\cite{hegarty99,loiko01,roessler03,chen03} and deserving further
studies.

\section*{ACKNOWLEDGEMENTS}
We want to thank J.\ Mulet and S.\ Balle for many discussions and
the fruitful collaboration on the subject and especially J.\ Mulet
for the possibility of using his programs for the theoretical
investigations. Some of the experimental data presented were
obtained in collaboration with  M.~Weinkath and C.~Engler in our
lab. We are also grateful to K.~Panajotov, J.~Danckaert and
G.~Verschaffelt for the collaboration on strain-induced
anisotropies and many discussions. The collaboration between the
groups was possible due to travel grants by the Deutsche
Akademische Austauschdienst and the COST Action 268. The
experimental work was supported by the Deutsche
Forschungsgemeinschaft. Finally, we gratefully acknowledge the
support and encouragement of W.\ Lange in our work.


\end{document}